\documentclass[preprint,1pt]{elsarticle}  

\usepackage{amssymb}
\usepackage{amsmath}
\usepackage{booktabs}
\usepackage[vcentermath]{youngtab}
\usepackage[dvipsnames]{xcolor}
\usepackage{bbm}
\usepackage{dsfont}
\usepackage{graphicx}
\usepackage[colorlinks=true, allcolors=blue]{hyperref}
\usepackage[vcentermath]{youngtab}
\usepackage{tikz}
\usepackage{todonotes}

\definecolor{blueca}{rgb}{0.0, 0.24, 0.82}
\newcommand{\su}[1]{\ensuremath{\mathrm{SU}(#1)}}

\newcommand{\Eref}[1]{Eq.~\eqref{#1}}
\newcommand{\tref}[1]{Tab.~(\ref{#1})}
\newcommand{\fref}[1]{Fig.~(\ref{#1})}
\DeclareMathOperator{\per}{\mathrm{Per}}
\newcommand{\imm}[1]{\ensuremath{\mathrm{Imm}^{\Yboxdim{5pt}\yng(#1,1)}}}

\usepackage[english]{babel}
\usepackage[utf8x]{inputenc}
\usepackage[T1]{fontenc}

\def\beq{\begin{equation}}
\def\eeq{\end{equation}}
\def\beqa{\begin{eqnarray}}
\def\eeqa{\end{eqnarray}}

\newcommand{\ket}[1]{\vert #1 \rangle}
\newcommand{\bra}[1]{\langle #1 \vert}

\newcommand{\complexityLabel}{\mathbb{P}}
\newcommand{\pla}{\ensuremath{\alpha}}
\newcommand{\plb}{\ensuremath{\beta}}
\newcommand{\plc}{\ensuremath{\gamma}}
\newcommand{\ca}{\ensuremath{\mathcal{A}}}
\newcommand{\cb}{\ensuremath{\mathcal{B}}}
\newcommand{\cc}{\ensuremath{\mathcal{C}}}
\newcommand{\wf}{\ensuremath{w}}
\newcommand{\Wf}{\ensuremath{\mathcal{W}}}
\newcommand{\outx}{\ensuremath{\eta}} 
\newcommand{\outy}{\ensuremath{\rho}} 
\newcommand{\outa}{\ensuremath{p}} 
\newcommand{\outb}{\ensuremath{q}} 
\newcommand{\frm}{\ensuremath{\mu}} 
\newcommand{\frM}{\ensuremath{\varepsilon}} 

\journal{Physics Letters A}

\begin{document}

\begin{frontmatter}



\title{Sum rules in multiphoton coincidence rates }

 \author[label1]{David Amaro Alcal\'{a}\corref{david}}
 \address[label1]{Instituto de F\'isica, Universidad Nacional
  Aut\'onoma de M\'exico, M\'exico D.F. 01000, M\'exico}
  \cortext[david]{Corresponding author}
  \ead{dav1494@ciencias.unam.mx}
  \author[label2]{Dylan Spivak}
   \address[label2]{{Department of Mathematical Sciences, Lakehead University, 955 Oliver Road, Thunder Bay, ON, Canada}}
  \author[label3]{Hubert de Guise}
 \address[label3]{Department of Physics, Lakehead University, 955 Oliver Road, Thunder Bay, ON, Canada}


\begin{abstract}
We show that sums of carefully chosen coincidence rates in a multiphoton interferometry experiment can be simplified by replacing the original unitary scattering matrix with a coset matrix containing $0$s.  The number and placement of these $0$s reduces the complexity of each term in the sum without affecting the original sum of rates.   In particular, the evaluation of sums of modulus squared of permanents is shown to turn in some cases into a sum of modulus squared of determinants.  The sums of rates are shown to be equivalent to the removal of some optical elements in the interferometer.
\end{abstract}


\end{frontmatter}

\section{Introduction}
\label{sec: intro}
The objective of this Letter is to highlight 
reductions in the computational complexity
of certain sums of 
coincidence rates 
for photons scattered in a passive optical network. 
The mathematics behind the result depends on orthogonality of subgroup
functions as will be shown in Sec.~{\ref{sec:general}}, but we also present our results in the context of interferometry, 
and discuss in particular how
the summation of specific rates could be obtained using a simpler interferometer where some
optical elements are removed.
\par
Although the results depend critically on eliminating a unitary \emph{submatrix}
of the scattering matrix $U$, the unitarity of $U$ itself does not enter in our arguments.  Thus
we envisage to use sums of rates and the ensuing simplifications to place constraints on the reconstruction
of matrices describing passive optical networks or the proper functioning of such devices.  
Another 
possible application is to use this technique in the context of certification, 
that is, testing the correctness of the output of an optical network
using a classical computer with reasonable resources.  These applications will be developed in 
future work.
\par

Our results are motivated in part by, but not restricted to the BosonSampling 
{\cite{aaronson2011computational}} problem, 
where permanents of submatrices of a unitary matrix are connected with
coincidence rates of fully indistinguishable photons.  

\begin{figure}[h!]
    \centering
    \includegraphics[height=3cm]{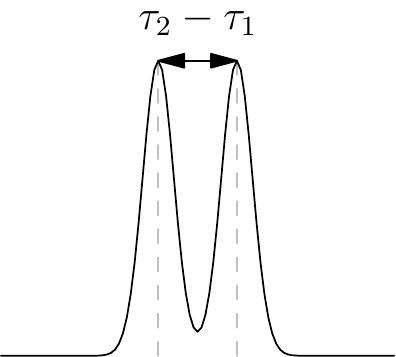}
    \caption{Two partially overlapping Gaussian pulses, separated by a time delay $\tau_2-\tau_1$.}
    \label{fig:partialoverlap}
\end{figure}

Indeed we need not here assume exactly indistinguishability: partial distinguishability between photons $i$ and $j$, is modelled by
the partial overlap of Gaussian wave packets describing these photons; 
as illustrated in \fref{fig:partialoverlap}, this overlap results from a 
time-delay $\tau_{i}-\tau_j$ 
between these wave packets and is an adjustable parameter \cite{tan20133,tillmann2015generalized}.
Other parametrizations for partial distinguishability for photons \cite{shchesnovich2015partial,tichy2015sampling} are possible.

When two photons enter a 2-channel interferometer and exactly overlap, the probability of detecting the two photons in different detectors
is given the modulus squared of the permanent of the matrix $U$ describing
the linear interferometer. The specific choice of a 50/50 beam splitter leads to 
no probability of getting one photon in each detector,
as  demonstrated in spectacular fashion by Hong, Ou, and Mandel \cite{hong1987measurement}.

The appearance of the modulus squared of a permanent is a generic feature of coincidence rates for fully
indistinguishable bosons (for fermions, a determinant would replace the permanent)
\cite{scheel2004permanents,lim2005generalized,spivak2017immanants}, and the computational
complexity of this permanent is at the core of the 
BosonSampling paradigm, where a dilute
collection of $n$ non-interacting bosons scatter inside an $m$-channel optical network
with $m\gg n$.   

Our results show that certain sums of rates
 - \textit{i.e.} sums of moduli squared of immanants - computed with the original scattering matrix $U$ 
are equal to the same sums if $U$
is replaced by a simpler \emph{coset matrix}
$\bar U$ containing strategically placed zeroes.  The number of zeroes and their placement depends in general on type of sums. 

We show explicitly for $n=3$ and $n=4$ there exists a sum that has the same value if 
$U$ is replaced by a coset matrix $\bar U$ of the Hessenberg type, a special
kind of ``almost upper (or lower) triangular'' matrix where $\bar{U}_{k,k+m}=0$ for $m=2,\ldots n-k$.  The permanent of 
a Hessenberg matrix 
is actually the \emph{determinant} of 
$T(\bar U)$, where $T(\bar U)$ is obtained from $\bar U$ by
replacing $U_{k,k+1}$ by $-U_{k,k+1}$ \cite{Hessenberg}.   
As a result, the complexity of certain sums of rates
is considerably simplified: general permanents evaluate by Ryser's algorithm in 
$\mathcal{O}(2^{n-1} n^2)$ 
operations whereas
determinants efficiently evaluate in $\mathcal{O}(n^3)$ operations; this is 
exemplified in \Eref{eq:sumpermassumdet},
where we give the sum of rates for  three indistinguishable photons. 
Moreover, some of the rates needed in the sums may contains 
multiple
photon counts per channel. Thus, the submatrix required to compute
the rate have rank less than $n$.  In some cases, one can revert to 
algorithms that allow for faster permanent computation~\cite{barvinok} of low rank matrices.
We also discuss how this is generalizable to any $n$.

When particles are partially distinguishable, the rates are expressed as a sum of moduli squared of \emph{immanants} 
\cite{littlewood,tan20133,tillmann2015generalized,wu2018coincidence,khalid2018permutational}. 
In the simple
case of two partially distinguishable photons, the sum contains terms proportional to
(modulus squared of the) permanent and the determinant of the scattering matrix, 
these being special types of immanants.  For the more interesting case of three input photons, 
one requires immanants of 
$3\times 3$ matrices: they are discussed
in {\ref{sec:3x3imm}}.  

Immanants of Hessenberg matrices 
can also be simplified as an easy corollary of results given by {\cite{Hessenberg}}:
the computational complexity of some immanants are also in the complexity class \#$\complexityLabel$
{\cite{burgisser2000generalcomputational,burgisser2000computational,hartmann1985complexity,brylinski2003complexity}},
but
the immanant of a Hessenberg matrix $\bar U$, associated with partition of shape $\{\lambda\}$, can be computed instead using the immanant associated with the conjugate shape $\{\lambda^*\}$ of a transformed Hessenberg matrix $T(\bar U)$. Thus certain sums of rates
for three or more partially distinguishable photons
are also simplified since we can choose to compute the simpler of the $\{\lambda\}$ or $\{\lambda^*\}$ immanants, although of course the savings are limited when $n$ is small. 
Ref. {\cite{burgisser2000computational}} provides an algorithm to evaluate the $\{\lambda\}$-immanant of 
an $n \times n$ matrix that has non-scalar complexity $\mathcal{O}(n^2 s_{\lambda} d_{\lambda})$, 
where $s_{\lambda}$ is the number standard tableaux of shape $\lambda$ and $d_{\lambda}$ is the number 
of semi-standard tableaux. An immanant and its conjugate will have $s_{\lambda}=s_{\lambda^*}$, 
so to determine which of the two immanants is harder to compute, we need to look at $d_{\lambda}$ 
and $d_{\lambda^*}$. In general, the partition with the fewer number of parts will have the greater 
number of semi-standard tableaux, and will thus be harder to compute. 
For example, the immanant corresponding to
\(\Yboxdim{5pt}\yng(4,1)\)
is more computationally expensive than its 
conjugate
\(\Yboxdim{5pt}\yng(2,1,1,1)\).


We 
provide here these simplifications for setups where $2$ photons interfere
inside a 3-channel device, and where $3$ photons interfere inside a $4$-channel device.  We explain how the simplified matrix $\bar U$ can 
be realized by removing elements in a unitary optical network. 
We also outline for the case of $n-1$ indistinguishable photons
entering a $n\times n$ network and the corresponding savings.

\section{$2$ photons in a $3$-channel interferometer}
\label{2photons3channel}

In this section we introduce the simplest case where savings by sum rules can
be achieved: $2$ photons in a $3$-channel interferometer.
\subsection{Connection with permanents and determinants}

The coincidence
rate for two partially distinguishable photons entering in
ports $2'$ and $3'$ of a $3$-port interferometer, see
Fig.~(\ref{fig:interferometers}a), 
and detected in ports $1$ and 
$3$, is given by 
\begin{align}
    R(23\to 13;\tau_{12})&= 
    e^{-\tau_{12}^2}\left(U_{12}^\dagger U_{33}^\dagger U_{32}U_{13}
    +U_{12}U_{33}U_{32}^\dagger U_{13}^\dagger \right) \nonumber 
                       \\&\quad+\vert U_{12}U_{33}\vert^2 + \vert U_{13}U_{32}\vert^2\, ,\label{eq:Rnotdiagonal}\\
    &= \textstyle{\frac{1}{2}}(1+e^{-\tau_{12}^2}) \vert
    \hbox{Per}(U_{23\to13})\vert^2  \nonumber 
  \\&\quad+\textstyle{\frac{1}{2}}(1-e^{-\tau_{12}^2}) \vert
  \hbox{Det}(U_{23\to13})\vert^2 \, .
    \label{eq:Rdiagonal}
\end{align}
with $\tau_{12}=\tau_1-\tau_2$.  An example of this type of calculation,
including the modelling of the detectors, is given in \ref{sec:ratecalculation}.

When the pulses exactly overlap, \emph{i.e.} when $\tau_2=\tau_1$, the rate collapses to the modulus
squared of the permanent of the submatrix $U_{23\to13}$, obtained from the original $3\times 3$ unitary
matrix $U$ by keeping rows $1,3$ and columns $2,3$:
\begin{align}
    U=\left(\begin{array}{ccc}
    U_{11}&U_{12}&U_{13}\\
    U_{21}&U_{22}&U_{23}\\
    U_{31}&U_{32}&U_{33}\end{array}\right)\, ,\quad 
    U_{23\to13}= \left(\begin{array}{cc}
     U_{12}&U_{13}\\
    U_{32}&U_{33}\end{array}\right)\, .
\end{align}

If the group elements $P_\sigma$ of $S_2$ are
$\{\mathds{1}, P_{13}\}$, and the action of $P_{\sigma}$ is defined by the permutation of columns in the polynomial $U_{i2}U_{j3}$ so that $P_{\sigma}U_{i2}U_{j3}=U_{\sigma(i)2}U_{\sigma(j)3}$ then
\begin{align}
     \hbox{Per}(U_{23\to13})&=
     \left(\mathds{1}+P_{13}\right)U_{12}U_{33}=
     U_{12}U_{33}+U_{32}U_{13}\, ,\\
     \hbox{Det}(U_{23\to13})&=
     \left(\mathds{1}-P_{13}\right)U_{12}U_{33}
     =U_{12}U_{33}-U_{32}U_{13}\, .
\end{align}
Note that, by construction, 
\begin{align}
    P_{13} \hbox{Per}(U_{23\to13})&=+ \hbox{Per}(U_{23\to13})\, ,\\
    P_{13}  \hbox{Det}(U_{23\to13})&=- \hbox{Det}(U_{23\to13})\, .
\end{align}
The permanent and the determinant 
examples of \emph{immanants}, which are polynomial functions in the entries of a matrix, constructed here using the representations of the permutation group $S_2$ for two photons.  These two
immanants come back to multiple of themselves under
\emph{any} permutation or rows or columns ($+1$ for permanents, $-1$ for determinants). 

\subsection{Summing over the outputs}

Suppose that, in addition to detecting photons in output ports
$1$ and $3$ as previously described, we also count output
photons at ports $2$ and $3$.  We obtain the rates for this process 
by copying \Eref{eq:Rdiagonal} with simple adjustment
of the appropriate indices:
\begin{align}
    R(23\to 23;\tau_{12})&= 
    {\textstyle\frac{1}{2}}(1+e^{-\tau_{12}^2}) \vert \hbox{Per}(U_{23\to23})\vert^2 \nonumber 
                                 \\&\quad+{\textstyle\frac{1}{2}}(1-e^{-\tau_{12}^2}) \vert \hbox{Det}(U_{23\to23})\vert^2 \, .
    \label{eq:R2323diagonal}
\end{align}

We now \emph{sum} $R(23\to 23;\tau_{12})+R(23\to 13;\tau_{12})$:
\begin{align}
  & \sum_{\outa=1,2} R(23\to\outa 3;\tau_{12}) 
   = {\textstyle\frac{1}{2}}(1+e^{-\frac{1}{2}\tau_{12}^2})
    \sum_{\outa=1,2}
    \vert \hbox{Per}(U_{23\to \outa 3})\vert^2
    \nonumber \\ &\qquad\qquad  
   + {\textstyle\frac{1}{2}}(1-e^{-\frac{1}{2}\tau_{12}^2})\sum_{\outa=1,2}
    \vert \hbox{Det}(U_{23\to \outa 3})\vert^2
\end{align}.

\begin{figure}[h!]
  \includegraphics[width=0.45\textwidth]{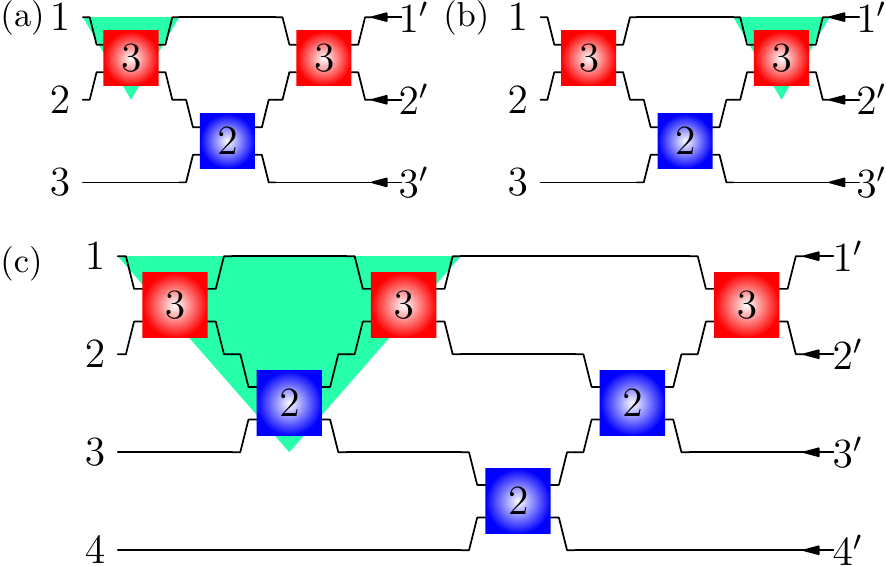}
\caption{
The numbers in each box represents the number of parameters in
the corresponding $\hbox{SU}(2)$ transformation. Numbers at
the sides label port number.
Input ports are on the right.
a)
Diagrammatic form of the decomposition of an $\hbox{SU}(3)$ matrix as a sequence of $\hbox{SU}(2)$ matrices.  The numbers in the boxes are the numbers of independent angles in each $\hbox{SU}(2)$ transformation.  
    When the rates at \emph{output} channels $(\outa 3)$ are summed over $\outa=1,2$, 
    the  $\hbox{SU}(2)$ transformation highlighted in green,
    ${\cal R}_{12}(\pla_3,\plb_3,\plc_3)$,
    can be replaced by the unit matrix, or the equivalent optical element can be removed, yielding a coset transformation 
    $\hbox{SU}(2)\backslash \hbox{SU}(3)$. See Eq.~\eqref{ec:su3-su2fact}.
b)
When rates at \emph{input} channels $(\outa 3)$ are summed over $\outa=1,2$, the
$\hbox{SU}(2)$ transformation highlighted in green, ${\cal
R}_{12}(\pla_1,\plb_1,\plc_1)$, can be replaced by the unit
matrix, or the equivalent optical element can be removed, yielding a coset
transformation $\hbox{SU}(3)/\hbox{SU}(2)$. See Eq.~\eqref{ec:su3-su2fact}.
c)
Diagrammatic representation of a $\hbox{SU}(4)$ matrix decomposed in
$\hbox{SU}(2)$ transformations. Green triangle highlights the $\hbox{SU}(3)$
submatrix, ${\cal R}_{123}$, see Eq.~\eqref{eq:su4factored}. 
}
    \label{fig:cosetmatrices}
\end{figure}

To highlight the (here elementary) simplification that occurs for this sum, we write the $3\times 3$ scattering
matrix $U$ in the form of a product \cite{rowe1999representations,simpleFact}, see Fig.~(\ref{fig:cosetmatrices}b)
\begin{align}
    U&= {\cal R}_{12}(\pla_1,\plb_1,\plc_1)
    \cdot \bar U\, ,\\
    &=
        \left(
\begin{array}{ccc}
 e^{-i \frac{1}{2}(\pla_1+ \plc_1)} \cos \left(\frac{\plb_1}{2}\right) & 
 -e^{-i \frac{1}{2}(\pla_1-\plc_1)}\sin \left(\frac{\plb_1}{2}\right) & 0 \\
e^{-i \frac{1}{2}(\plc_1-\pla_1)}\sin \left(\frac{\plb_1}{2}\right) & 
e^{i\frac{1}{2}(\pla_1+\plc_1)} 
\cos \left(\frac{\plb_1}{2}\right) & 0 \\
 0 & 0 & 1 \\
\end{array}
\right) \nonumber 
  \\&\qquad
\cdot \left(\begin{array}{ccc} 
    \bar{U}_{11}&\bar{U}_{12}&0\\
    \bar{U}_{21}&\bar{U}_{22}&\bar{U}_{23}\\
    \bar{U}_{31}&\bar{U}_{32}&\bar{U}_{33}
    \end{array}\right)\, ,
\end{align}
where the \emph{unitary} transformation 
${\cal R}_{12}(\pla_1,\plb_1,\plc_1)$ is an $\mathrm{SU}(2)$ transformation
mixing the first and second channels. 
Other factorizations into $\mathrm{SU}(2)$ or $\mathrm{U}(2)$ blocks are possible {\cite{reck1994experimental,murnaghan1962unitary,clements2016optimal,russell2017direct}} , but do not 
produce the easily identifiable coset structure required for the general $n\times n$ submatrix.  The algorithms
of {\cite{urias2010householder}} or {\cite{cabrera2010canonical}} can also be used to efficiently obtain a suitable coset factorization.

One can then easily verify, using the factorized form of $U$, that
the first two rows of $U$ are made to depend explicitly on the  parameters
$\pla_1,\plb_1,\plc_1$ so that each of
$\vert \hbox{Per}(U_{23\to23})\vert^2$, 
$\vert \hbox{Det}(U_{23\to23})\vert^2$, 
$\vert \hbox{Per}(U_{23\to13})\vert^2$ and 
$\vert \hbox{Det}(U_{23\to13})\vert^2$
individually depends on these parameters. However, the sums
\begin{align}
&\vert \hbox{Per}(U_{23\to23})\vert^2
+\vert \hbox{Per}(U_{23\to13})\vert^2\, , \\
\hbox{and }\quad &\vert \hbox{Det}(U_{23\to23})\vert^2 + 
\vert \hbox{Det}(U_{23\to13})\vert^2
\end{align}
are actually independent of 
$\pla_1,\plb_1,\plc_1$.  We denote this independence 
using the coset notation $\su{2}\backslash U$, 
and refer to $\bar U$ as a coset matrix. 
We will show
in detail the origin of this independence in Sec.~{\ref{sec:general}}.

We are therefore free to choose $\pla_1,\plb_1,\plc_1$ as we please: the simplest choice is to make
${\cal R}_{12}$ the unit matrix with $\pla_1=\plb_1=\plc_1=0$, so that we have an example of the core
result of this Letter:
\begin{align}
\sum_{\outa=1,2}\vert \hbox{Per}(U_{23\to \outa 3})\vert^2
&=\sum_{\outa=1,2} \vert \hbox{Per}(\bar{U}_{23\to \outa 3})\vert^2
\, \label{eq:savin1}, \\
  \sum_{\outa=1,2}\vert \hbox{Det}(U_{23\to \outa 3})\vert^2
&=\sum_{\outa=1,2}\vert \hbox{Det}(\bar{U}_{23\to \outa 3})\vert^2\, \label{eq:savin2}.
\end{align}
In particular we note that both
\begin{align}
  \hbox{Per}(U_{23\to13})&=
  \hbox{Per}\left(\begin{array}{cc} 
  \bar U_{12} & 0 \\
  \bar U_{32} & \bar U_{33}
  \end{array}\right)\\
  \hbox{Det}(U_{23\to13})&=
  \hbox{Det}\left(\begin{array}{cc} 
  \bar U_{12} & 0 \\
  \bar U_{32} & \bar U_{33}
  \end{array}\right)
\end{align}
trivially evaluate to $\bar{U}_{12}\bar{U}_{33}$.  

Another choice is $\pla_1=\plc_1=0$ but $\plb_1=\pi$ so the $\mathcal{R}_{12}$ matrix takes the form
\begin{align}
{\cal R}_{12}(0,\pi,0)=
\left(\begin{array}{ccc}
0&1&0\\
-1&0&0\\
0&0&1
\end{array}\right)    
\end{align}
which yields
\begin{align}
    {\cal R}_{12}(0,\pi,0)\bar U
    =\left(\begin{array}{ccc}
    -\bar{U}_{21}&-\bar{U}_{22}&-\bar{U}_{23}\\
    \bar{U}_{12}&\bar{U}_{22}&0\\
    \bar{U}_{13}&\bar{U}_{23}&\bar{U}_{33}
    \end{array}\right)\, ,
\end{align}
showing that the results are essentially unchanged if the $0$ appears in on the 
second row of the last column.

If we now assume $U$ is unitary and, without loss of generality,
with determinant $+1$, the resulting $U$ is an $\hbox{SU}(3)$ 
transformation.  We can then realize the $\hbox{SU}(3)$ transformation
describing the inteferometer as a sequence 
of $\hbox{SU}(2)$ inteferometers mixing modes $(12)$, $(23)$ and 
then $(12)$ again \cite{simpleFact}:
\begin{equation}
  U={\cal R}_{12}(\pla_1,\plb_1,\plc_1){\cal
  R}_{23}(\pla_2,\plb_2,\pla_2){\cal
R}_{12}(\pla_3,\plb_3,\plc_3)\, .
  \label{ec:su3-su2fact}
\end{equation}
Summing the rates for $U$ then yields the same result as summing
the rates over a scattering matrix $\bar U$ describing an inteferometer
with the rightmost element
removed,  as illustrated
in Fig.~(\ref{fig:cosetmatrices}b). 

\subsection{Symmetry analysis}

One can understand the origin of the independence on the $\hbox{SU}(2)$ parameters
$\pla_1,\plb_1,\plc_1$ as follows.  
Define
\begin{align}
    \hat A^\dagger_k(\tau_m)
    &\equiv\int d\frm\, e^{i\frm\tau_m}
    \phi(\frm)\hat a_k^\dagger(\frm)\, ,\label{eq:bigAop} \\
    \hat C_{ij}&=A_i^\dagger(\tau_1)\hat A_j(\tau_1)+\hat A_i^\dagger(\tau_3)\hat A_j(\tau_3)\, ,
\end{align}
where $\phi(\frm)$ is the spectral profile of a pulse in channel $k$,
and construct the $\mathfrak{su}(2)$ subalgebra 
\begin{align}
    \hat C_{12}&\mapsto \hat L_+\, ,   \qquad  \hat C_{21}\mapsto \hat L_-\, ,\quad 
    \hat C_{11}-\hat C_{22}\mapsto 
    2\hat L_z\, . \label{eq:su2subalgebra}
\end{align}
One can easily verify that 2-photon output 
states of the type
\begin{align}
   \ket{\psi_{13}}_+&=\left(\hat A^\dagger_1(\tau_1)\hat A_3^\dagger(\tau_3)+  \hat A^\dagger_1(\tau_3)\hat A_3^\dagger(\tau_1)
   \right) \ket{0}\, , \label{eq:20su3}\\
     \ket{\psi_{13}}_-&=\left(\hat A^\dagger_1(\tau_1)\hat A_3^\dagger(\tau_3)-  \hat A^\dagger_1(\tau_3)\hat A_3^\dagger(\tau_1)
   \right) \ket{0} \label{eq:01su3}
    \end{align} 
are killed by $\hat C_{12}$ so the sets 
$\{\ket{\psi_{13}}_\pm, \hat C_{21}\ket{\psi_{13}}_\pm\}_\pm$ each span a 2-dimensional 
representation of this $\mathfrak{su}(2)$.
(Note that, 
when $\tau_3=\tau_1$, only the states  $\ket{\psi_{13}}_+$ and 
$\hat C_{21}\ket{\psi_{13}}_+$ survive.)

Therefore, by summing over detected states of the type 
$\hat A^\dagger_1(\tau_i)\hat A_3^\dagger(\tau_j)\ket{0}$ 
and $\hat A^\dagger_2(\tau_i)\hat A_3^\dagger(\tau_j)\ket{0}$, we are summing over complete
sets of $\mathfrak{su}(2)$ states, eliminating the dependence on the matrix 
${\cal R}_{12}(\pla_1,\plb_1,\plc_1)$.



\subsection{Summing over the inputs}

A similar conclusion is reached ---assuming both
input configurations with equal time delays, \emph{v.g.}, $\tau_2-\tau_3=\tau_1-\tau_3$
if we fix the output to ports $2,3$ but now
sum over the input channels $1,3$ and $2,3$.
In this case, the rates are computed using submatrices of the type
\begin{align}
   {U}_{\outa 3\to23} =
\begin{pmatrix}
   {U}_{2\outa} &   {U}_{23} \\
   {U}_{3\outa} &  {U}_{33}
\end{pmatrix}\, ,\qquad \outa=1,2\, .
\end{align}

We now factorize the scattering matrix $U$ as the product
\begin{align}
    U&= \tilde{U} \cdot {\cal R}_{12}(\pla_3,\plb_3,\plc_3)  \, ,\\
    &=
\left(\begin{array}{ccc} 
    \tilde{U}_{11}&\tilde{U}_{12}&\tilde{U}_{13}\\
    \tilde{U}_{21}&\tilde{U}_{22}&\tilde{U}_{23}\\
    0&\tilde{U}_{32}&\tilde{U}_{33}
    \end{array}\right)
  \\&\quad \cdot
        \left(
\begin{array}{ccc}
 e^{- \frac{i}{2}(\pla_3+ \plc_3)} \cos \left(\frac{\plb_3}{2}\right) & 
 -e^{- \frac{i}{2}(\pla_3-\plc_3)}\sin \left(\frac{\plb_3}{2}\right) & 0 \\
e^{- \frac{i}{2}(\plc_3-\pla_3)}\sin \left(\frac{\plb_3}{2}\right) & 
e^{\frac{i}{2}(\pla_3+\plc_3)} 
\cos \left(\frac{\outb_3}{2}\right) & 0 \\
 0 & 0 & 1 \\
\end{array}
\right)\, , \label{eq:su3/su2}
\end{align}
as illustrated in  Fig.~(\ref{fig:cosetmatrices}a).
Summing the rates over inputs $1,3$ and $2,3$, we find
the sum does not depend on $\pla_3,\plb_3,\plc_3$,
so we can choose ${\cal R}_{12}$ to be the unit matrix and use
$\tilde{U}$, with the leftmost element removed, as
illustrated in Fig.~(\ref{fig:cosetmatrices}a).

\section{$3$ photons in a $4$-channel interferometer:  $\hbox{SU}(3)\backslash U$}
\label{sec:su4}

In this section we discuss the savings resulting from sum rules corresponding to the
extension of the previous case: $3$ photons in a $4$-channel interferometer.
We present this case because it illustrates all the features present 
in configurations with more than $2$ photons.

\subsection{Summing over outputs}

We consider without loss of generality the situation when $3$ photons access to the
interferometer by channels $1'$, $2'$, and $4'$, while detectors 
are put at output channels $2$, $3$ and $4$.
We will sum over processes where
one of the three photons is always counted 
in channel $4$.

\begin{figure}[h!]
\includegraphics[width=0.475\textwidth]{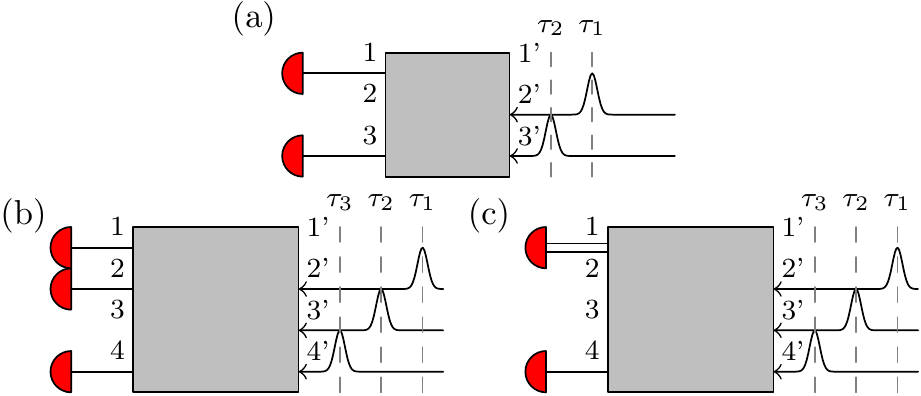}
 \caption{
   a) Example of a process involved in the detection of two photons entering
 in ports $2$ and $3$.  Input photons in modes $2$ and $3$ are counted in 
 detectors $1$, and $3$.
 Bottom. Examples of processes entering in the detection of three photons entering
   in ports $2,3$ and $4$.  b) input photons in modes $2,3,4$ are counted in 
   detectors $1$, $2$ and $4$.  c) input photons in modes $2,3,4$ are counted in 
 detectors $1$  (2 counts) and in detector 4.
}
    \label{fig:interferometers}
\end{figure}

First, suppose photons are counted one each in detectors $1,2$ and $4$, as in Fig.~(\ref{fig:interferometers}b). 
Computation of the rates now involves immanants of the $3\times 3$ submatrix  
\begin{align}
U_{234\to 124}=\left(\begin{array}{ccc}
U_{12}&U_{13}&U_{14}\\
U_{22}&U_{23}&U_{24}\\
U_{42}&U_{43}&U_{44}
\end{array}\right)\,  \label{eq:full234to123}
\end{align}
obtained by keeping columns $2,3,4$ and rows $1,2,4$.  Similarly if photons are counted 
in detectors $1,3,4$ we keep now rows $1,3,4$, and if they are counted one each
in detectors $2,3,4$ we keep rows $2,3,4$ of the submatrix.

If one photon is counted in detector $4$ but two are counted in detector $2$, we need to duplicate row $2$ in the submatrix:
\begin{align}
U_{234\to 224}=\left(\begin{array}{ccc}
U_{22}&U_{23}&U_{24}\\
U_{22}&U_{23}&U_{24}\\
U_{42}&U_{43}&U_{44}
\end{array}\right)\, ,
\end{align}
and similarly appropriately duplicate rows when two photons are counted in detector $3$ and one in detector $4$, and two
are counted in detector $2$ and one in detector $4$.  
This feature of summing rates with multiple
photons in one port is not present in the first case Sec.~\ref{2photons3channel}, 
and also
not present in BosonSampling, where the probability of having multiple photons in a single output
detector is kept low by diluting $n\ll m$.
\par
If we now sum the rates associated with this input setup we find,
irrespective of the relative input delays , that the sums 
are all identical to those obtained by using the appropriate submatrices of the simpler matrix
\begin{align}
\bar U= \left(\begin{array}{cccc}
\bar{U}_{11}&\bar{U}_{12}&0&0\\
\bar{U}_{21}&\bar{U}_{22}&\bar{U}_{23}&0\\
\bar{U}_{31}&\bar{U}_{32}&\bar{U}_{33}&\bar{U}_{34}\\
\bar{U}_{41}&\bar{U}_{42}&\bar{U}_{43}&\bar{U}_{44}
\end{array}\right)\, . \label{eq:su3cosetsu4}
\end{align}
In other words, we write the full scattering matrix 
\begin{align}
    U={\cal R}_{123}\cdot \bar U\, ,\quad 
     {\cal R}_{123}=\left(\begin{array}{cccc}
    *&*&*&0\\
    *&*&*&0\\
    *&*&*&0\\
    0&0&0&1\end{array}\right) \label{eq:su4factored}
\end{align}
with ${\cal R}_{123}$ an SU$(3)$ matrix, depending on $8$ parameters, 
mixing only modes { $1$, $2$ and $3$}.  With this factorization one shows that
the \emph{sum} of rates is independent of the $8$ parameters of ${\cal R}_{123}$. 

Assuming the transformation $U$ is unitary with determinant $+1$,
we can realize this transformation as an SU$(4)$ 
interferometer decomposed in a sequence of $\hbox{SU}(2)$ 
transformations as in
\cite{simpleFact}.   The independence of
the sum on the $8$ parameters of ${\cal R}_{123}$ is equivalent to removing three optical elements in
the system, as illustrated in Fig.~(\ref{fig:cosetmatrices}c).

From \Eref{eq:su3cosetsu4} it follows that the submatrices of $\bar U$ are of the form
\begin{align}
\bar{U}_{234\to 124}=\left(\begin{array}{ccc}
\bar{U}_{12}&0&0\\
\bar{U}_{22}&\bar{U}_{23}&0\\
\bar{U}_{42}&\bar{U}_{43}&\bar{U}_{44}
\end{array}\right)\, . \label{eq:U234to123}
\end{align}
Two other examples of submatrices needed are 
\begin{align}
\bar{U}_{234\to 134}&=\left(\begin{array}{ccc}
\bar{U}_{12}&\bar{U}_{13}&0\\
\bar{U}_{32}&\bar{U}_{33}&\bar{U}_{34}\\
\bar{U}_{42}&\bar{U}_{43}&\bar{U}_{44}
\end{array}\right)\, , 
 \label{eq:U234to134}\\ 
\bar{U}_{234\to 224}&=
\left(\begin{array}{ccc}
\bar{U}_{22}&\bar{U}_{23}&0\\
\bar{U}_{22}&\bar{U}_{23}&0\\
\bar{U}_{42}&\bar{U}_{43}&\bar{U}_{44}
\end{array}\right)\, . \label{eq:U234to144}
\end{align}

The matrix $\bar U$ and the submatrices of
\eqref{eq:U234to123}, \eqref{eq:U234to134}
and \eqref{eq:U234to144} are of the \emph{Hessenberg} type, 
\emph{i.e.} matrices where $\bar{U}_{i,i+k}=0$ for $k\ge 2$.   These have the following important property: the computation of the 
permanent of such matrices can be mapped to the computation of the determinant of a matrix $T(U)$, obtained from $U$
by changing the entries $U_{i,i+1}$ to their negatives \cite{Hessenberg}.  In other words, we have for instance
\begin{align}
\hbox{Per}(\bar{U}_{234\to 134}) &= \hbox{Det}(T(\bar U)_{234\to 134})\, ,\\ 
T(\bar U)_{234\to 134}&=\left(\begin{array}{ccc}
\bar{U}_{12}&-\bar{U}_{13}&0\\
\bar{U}_{32}&\bar{U}_{33}&-\bar{U}_{34}\\
\bar{U}_{42}&\bar{U}_{43}&\bar{U}_{44}
\end{array}\right)\, , \label{eq:per_maps_to_det}
\end{align}
and similarly for the other matrices of the Hessenberg type.   In particular, the matrix $\bar{U}_{234\to 124}$ is triangular so we have
$\hbox{Per}(\bar{U}_{234\to 123})=\hbox{Det}(\bar{U}_{234\to 123})=\bar{U}_{12}\bar{U}_{23}\bar{U}_{44}$.

Thus, for instance, if all three photons are coincident at input, the sum of output
rates is a sum of permanents of submatrices:
\begin{align}
    &\sum_{\outa=1}^2 \sum_{\outb=\outa+1}^3 \vert \hbox{Per}
    (U_{234\to \outa\outb 4})\vert^2
    +{\textstyle\frac{1}{2}}\sum_{\outa=1}^3 
    \vert \hbox{Per}
    (U_{234\to \outa\outa 4})\vert^2 \nonumber \\
    &= \sum_{\outa=1}^2 \sum_{\outb=\outa+1}^3 \vert \hbox{Per}
    (\bar U_{234\to \outa\outb 4})\vert^2
    +{\textstyle\frac{1}{2}}\sum_{\outa=1}^3 
    \vert \hbox{Per}
    (\bar U_{234\to \outa\outa 4})\vert^2\, ,\label{eq:su3sumofrates}\\
 &= \sum_{\outa=1}^2 \sum_{\outb=\outa+1}^3 \vert \hbox{Det}
    (T(\bar U)_{234\to \outa\outb 4})\vert^2
    +{\textstyle\frac{1}{2}}\sum_{\outa=1}^3 
    \vert \hbox{Per}
    (\bar U_{234\to \outa\outa 4})\vert^2\, ,\label{eq:sumpermassumdet}
\end{align}
Note that an extra factor $\frac{1}{2}$ multiplies those terms describing the
detection of two identical
photons in the same detector since a state containing two identical particles has an extra 
$\sqrt{2!}$ denominator factor for proper normalization.
The origin of this extra factor is discussed at some greater length in Sec.~{\ref{sec:general}}.

\subsection{Partially indistinguishable wave packets and Immanants}

Although more complicated to generalize, the situation is more interesting when not all three 
photons exactly overlap.
There are now three possible types of rates, associated with the three possible Young diagrams labelling the irreps of $S_3$.  
They are $\Yboxdim{5pt}\yng(3)$, $\Yboxdim{5pt}\yng(1,1,1)$ and 
$\Yboxdim{5pt}\yng(2,1)$.  The first type corresponds to the fully
symmetric representation: if the three input photons are fully indistinguishable, the coincidence rates are \emph{only} a function 
of permanents of a $3\times 3$ submatrix. 
 If two of the photons are indistinguishable,
the rates now depend not only on the permanent of a submatrix but also on some immanants of the type $\Yboxdim{5pt}\yng(2,1)$
of this submatrix.  If all three photons are partially indistinguishable, the rates also include 
contributions from the determinant of a $3\times 3$ matrix
in addition to the previous immanants.

Immanants generalize permanents and determinants, and additional details on
immanants of a $3\times 3$ matrix can be found in \ref{sec:3x3imm}.  

Our formalism also requires the counting two photons in any one of the detectors.  We can accommodate this by using a $3\times 3$ submatrix constructed
from the full scattering matrix by duplicating the appropriate column or row of $U$.
The immanants of this submatrix are then computed in the usual way (of course
in this case any determinant is automatically $0$).

We discuss here the case
where $\tau_1=\tau_2 \ne \tau_3$, and we assume
for simplicity that $U$ is unitary with determinant $+1$.  
In this case the 3-photon input states belong to the 
irreps $(3,0,0)$ (or $\Yboxdim{5pt}\yng(3)$) of 
$\hbox{SU}(4)$, or to the irrep $(1,1,0)$ (or $\Yboxdim{5pt}\yng(2,1)$) of 
$\hbox{SU}(4)$. Rates are no longer given by the permanent of a submatrix, but must also include
immanants associated with $\Yboxdim{5pt}\yng(2,1)$ partition of the permutation group $S_3$ of 
the three photons, see Eq.~\eqref{eq-immanant-3}.  

For instance: 
\begin{align}
   & R(234\to \outa\outa 4;\tau_{13}) 
    = {\textstyle\frac{1}{3}}(1+2e^{-\tau_{13}^2})
    |\per (U_{234\to\outa\outa 4})|^2\nonumber \\
&+     {\textstyle\frac{2}{3}}(1-e^{-\tau_{13}^2})
    |\imm{2}(U_{234\to\outa\outa 4})|^2
    \, ,\label{234topp4}\\
    &R(234\to \outa\outb 4;\tau_{13}) = 
    \vert \ca\vert^2 + \vert \cb\vert^2 + \vert \cc\vert^2 \nonumber \\
    &+e^{-\tau_{13}^2}
    \left[(\ca+\cb)^*\cc+ (\cb+\cc)^* + (\cc+\ca)^* \cb\right], \label{234topq4}
\end{align}
where the functions $\ca$, $\cb$, and $\cc$ are related to
immanants by
\begin{align}
  \ca &= {\textstyle\frac{1}{3}}(\per (U_{234\to\outa\outb 4}) - \imm{2}(U_{243\to\outa\outb 4}) 
  \nonumber \\&\qquad- \imm{2}(U_{324\to\outa\outb 4})  + \imm{2}(U_{342\to\outa\outb 4})),  \\
  \cb &= {\textstyle\frac{1}{3}}(\per (U_{234\to\outa\outb 4}) - \imm{2}(U_{234\to\outa\outb 4}) 
 \nonumber 
  \\&\quad  + \imm{2}(U_{243\to\outa\outb 4}) - \imm{2}(U_{324\to\outa\outb 4}) 
  \nonumber 
  \\&\quad - \imm{2}(U_{342\to\outa\outb 4})),  \\
  \cc &= {\textstyle\frac{1}{3}}(\per( U_{234\to\outa\outb 4}) + \imm{2}(U_{234\to\outa\outb 4})) 
  \nonumber 
  \\&\quad+ \imm{2} (U_{324\to\outa\outb 4})),
\end{align}
with $\tau_{13}=\tau_1-\tau_3$.   The notation
$\imm{2}(U_{ijk\to\outa\outb 4})$ indicates that the immanant is calculated using the matrix $U_{ijk\to\outa\outb 4}$
where the columns of $U_{234\to\outa\outb 4}$ are permuted to the order $ijk$.   
Both Eqs.(\ref{234topp4}) and (\ref{234topq4}) 
correctly collapse to a single permanent when $\tau_{13}=0$.

With the appearances of immanants, one can ask if
simplifications similar to those of \Eref{eq:sumpermassumdet} occur.  
Indeed one can show that the immanant $\hbox{Imm}^{\{\lambda\}}(\bar U) $ of
a Hessenberg matrix $\bar{U}$ maps to the calculation of the immanant $\hbox{Imm}^{\{\lambda\}^*}(T(\bar{U}))$ where 
$\{\lambda\}^*$ is the partition conjugate to $\{\lambda\}$.
In the specific case of immanants of the type $\Yboxdim{7pt}\yng(2,1)$ needed for our $3$-photon problem, 
there is no associated savings as the partition $\Yboxdim{7pt}\yng(2,1)$ is self-conjugate.

\section{$n-1$  photons in a $n$-channel interferometer}
\label{sec:general}
In this section we discuss the mathematical origin of the simplifications in sums presented in previous sections, Eqs.~{\eqref{eq:savin1}},
{\eqref{eq:savin2}}, and {\eqref{eq:sumpermassumdet}}. First,
we reconsider the case in Sec.~{\ref{2photons3channel}} in terms of group functions~ {\cite{spivak2017immanants,kostan1995immanant}}.
We then proceed to extend the result for $n$ indistinguishable photons.
In this section we assume that the transformation $U$ is unitary.
\par


\subsection{$2$ photons in a $3$-channel interferometer in terms of group functions}
\label{sec-group-funct}
Irreducible representations of the unitary groups, like
irreducible representations of the permutation group, 
are also labelled by Young diagrams
\cite{lichtenberg2012unitary}. In this
notation, a single photon state in a $3$-channel interferometer 
will transform by representation $(1,0)$ of $\mathrm{SU}(3)$, whereas
two-photon states will transform by the representation $(1,0)\otimes (1,0)$.  This representation is reducible, and the reduction \cite{o1982closed, wesslen2008geometric} indeed often uses Young diagrams
as a convenient calculational device \cite{lichtenberg2012unitary,lopez1990young} :
\begin{align}
    \begin{array}{ccccccc}
    (1,0)&\otimes& (1,0)&= &(2,0)&\oplus &(0,1) \\
    \Yboxdim{7pt}\yng(1)&\otimes &\Yboxdim{7pt}\yng(1)&=& \Yboxdim{7pt}\yng(2)&\oplus
    &\Yboxdim{7pt} \yng(1,1)
    \end{array} \, .
\end{align}
The $\hbox{SU}(3)$ irrep $(a,b)$ has dimension $\frac{1}{2}(a+1)(b+1)(a+b+2)$. 
The representation $(2,0)$ or $\Yboxdim{7pt}\yng(2)$ is therefore $6$-dimensional; it contains the symmetric states
$\{\ket{100}\ket{100}$,$\frac{1}{\sqrt{2}}\bigl(\ket{100}\ket{010}$+
$\ket{010}\ket{100}\bigr)\ldots \}$, immediately generalizing to $\hbox{SU}(3)$ the well known spin-triplet states. 
The
$3$-dimensional representation $(0,1)$ or $\Yboxdim{7pt}\yng(1,1)$ contains the antisymmetric states 
$\{\frac{1}{\sqrt{2}}\left(\ket{100}\ket{010}-\ket{010}\ket{100}\right)\ldots \}$, again
generalizing to $\hbox{SU}(3)$ the well-known antisymmetric $\hbox{SU}(2)$ singlet.

Immanants are quite generally related to group functions \cite{Dfuntionsandimmanants}: 
\begin{align}
    \hbox{Per}\left(\begin{array}{cc}
    U_{2\outa}&U_{23}\\
    U_{3\outa}&U_{33}\end{array}\right)=
    D^{\Yboxdim{5pt}\yng(2)}_{(23);(\outa 3)}(U)\, , \\
     \hbox{Det}\left(\begin{array}{cc}
    U_{2\outa}&U_{23}\\
    U_{3\outa}&U_{33}\end{array}\right)=
    D^{\Yboxdim{5pt}\yng(1,1)}_{(23);(\outa 3)}(U)\, ,
\end{align}
where $(\outa 3)$ denotes a state where photons 
enter in input channel $\outa=1,2$ and in channel $3$, so that for instance
$(13)$ for photons entering in input channels $1$ and $3$.

In this notation, we therefore have, for the summation over the inputs with
fixed output channels, 
\begin{align}
    \sum_{\outa=1,2}
    \vert D^{\Yboxdim{5pt}\yng(2)}_{(23);(\outa 3)}(U)\vert^2 =
    \sum_{\outa=1,2}
    \vert D^{\Yboxdim{5pt}\yng(2)}_{(23);(\outa 3)}(\bar U)\vert^2\, , \\
     \sum_{\outa=1,2}
    \vert D^{\Yboxdim{5pt}\yng(1,1)}_{(23);(\outa 3)}(U)\vert^2 =
    \sum_{\outa=1,2}
    \vert D^{\Yboxdim{5pt}\yng(1,1)}_{(23);(\outa 3)}(\bar U)\vert^2
\end{align}
where the $(13)$ and $(23)$ states span an $\hbox{SU}(2)$ subrepresentation of $\hbox{SU}(3)$ inside the $\Yboxdim{7pt}\yng(2)$ 
representation, and a subrepresentation of $\hbox{SU}(3)$ in the 
$\Yboxdim{7pt}\yng(1,1)$ 
representation.

The corresponding sums over outputs with fixed inputs are simply
\begin{align}
    \sum_{\outa=1,2}
    \vert D^{\Yboxdim{5pt}\yng(2)}_{(\outa 3);(23)}(U)\vert^2 =
    \sum_{\outa=1,2}
    \vert D^{\Yboxdim{5pt}\yng(2)}_{(\outa 3);(23)}(\bar U)\vert^2\, , \\
     \sum_{\outa=1,2}
    \vert D^{\Yboxdim{5pt}\yng(1,1)}_{(\outa 3);(23)}(U)\vert^2 =
    \sum_{\outa=1,2}
    \vert D^{\Yboxdim{5pt}\yng(1,1)}_{(\outa 3);(2 3)}(\bar U)\vert^2\, .
\end{align}

We can show this explicitly for the permanent as follows.  Let us denote by $\vert (2,0)\outa_1\outa_2;I\rangle$ a basis states for the $(2,0)$ (or $\Yboxdim{5pt}\yng(2)$) irrep of $\mathfrak{su}(3)$, with $I$ the $\mathfrak{su}(2)$ label form states transforming by the irrep $I$ of the ${\cal R}_{12}$ subgroup. Here 
$(\outa_1,\outa_2)=(1,1)$ denotes two photons in mode $1$, while $(\outa_1,\outa_2)=(1,3)$ denote
one photon in mode $1$ and one in mode $3$.

Then \cite{Dfuntionsandimmanants}:
\begin{align}
\hbox{Per}(U_{23\to \outb 3})= \langle (2,0) \outb 3 \vert U \vert (2,0) 23\rangle
\end{align}
With this notation 
\begin{align}
&\sum_{\outb=1,2} \vert \hbox{Per}(U_{23\to \outb 3}) \vert^2\nonumber \\
&= \sum_{\outb=1,2}\langle (2,0)\outb 3\vert U \vert (2,0) 23\rangle \langle (2,0)\outb 3 \vert U \vert (2,0) 23\rangle^*\, .
\end{align}
At this point, we split the transformation 
$U={\cal R}_{12}(\omega_1)\bar U$, with $\omega_1 =
(\pla_1,\plb_1,\plc_1)$ parametrizing an $\mathrm{SU}(2)$ transformation:
\begin{align}
&\sum_{\outb=1,2} \vert \hbox{Per}(U_{23\to \outb 3}) \vert^2 \nonumber \\
&=\sum_{\outb, \gamma,\gamma'=1,2}
\langle (2,0)\outb 3\vert {\cal R}_{12}(\omega_1)
\vert (2,0)\gamma 3 \rangle 
\langle (2,0)\gamma 3 \vert \bar U  \vert (2,0) 23\rangle
\nonumber \\
&\times \langle (2,0)\outb 3 \vert {\cal R}_{12}(\omega_1)
\vert (2,0)\gamma' 3\rangle^* 
\langle (2,0)\gamma' 3\vert \bar U  \vert (2,0) 23\rangle
^*\, ,
\end{align}
and explicitly use the $\mathrm{SU}(2)$ $D$-function for the ${\cal R}_{12}(\omega_1)$ transformation
\begin{align}
&\sum_{\outb=1,2} \vert \hbox{Per}(U_{23\to \outb 3}) \vert^2
=\sum_{\gamma\gamma'}
\left(\sum_{\outb} D^{1/2}_{\Wf(\outb) \wf(\gamma)}(\omega_1)
\left(D^{1/2}_{\Wf(\outb) \wf(\gamma')}(\omega_1)\right)^*\right) \nonumber \\
&\qquad \times 
\langle (2,0)\gamma 3 \vert \bar U  \vert (2,0) 23 \rangle^* 
\langle (2,0)\gamma' 3 \vert \bar U  \vert (2,0) 23\rangle^*\, ,
\end{align}
where
\begin{align}
\Wf(\outb)=\left\{\begin{array}{cc}
+\frac{1}{2}&\hbox{if } \outb=1;\\
-\frac{1}{2}&\hbox{if } \outb=2, \end{array}\right.
\end{align}
and $\wf(\gamma)$ likewise defined so it takes the values $\pm \frac{1}{2}$.  
The sums of $D$-functions with the same angle satisfy:
\begin{align}
\sum_{\Wf(\outb)} D^{1/2}_{\Wf(\outb) \wf(\gamma)}(\omega_1)
\left(D^{1/2}_{\Wf(\outb) \wf(\gamma')}(\omega_1)\right)^* =\delta_{\gamma\gamma'}
\end{align}
so that the sum collapses to
\begin{align}
&\sum_{\outb=1,2} \vert \hbox{Per}(U_{23\to \outb 3}) \vert^2\nonumber \\
&=\sum_{\gamma}
\langle (2,0)\gamma 3 \vert \bar{U}  \vert (2,0) 23\rangle 
\langle (2,0)\gamma 3 1\vert \bar{U}  \vert (2,0) 23\rangle^*\, , \\
&=\sum_{\gamma =1,2} \vert \hbox{Per}(\bar{U}_{23\to \gamma 3}) \vert^2\, ,
\end{align}
which shows the first part of the result. 

The results for the determinant follows the same steps, but with the replacement of the irrep $(2,0)$ (or $\Yboxdim{5pt}\yng(2)$) by $(0,1)$ or $(2,0)$ (or $\Yboxdim{5pt}\yng(1,1)$), and the identification 
\begin{align}
    \hbox{Det}(U_{23\to\outb 3})=\langle (0,1) \outb 3 \vert U \vert (0,1) 23\rangle\, .
\end{align}

\subsection{Generalization for $n-1$  photons in a $n$-channel interferometer}
We extend the ideas presented in Sec.~\ref{sec-group-funct} to the general case of 
$n-1$ indistinguishable photons.
We consider the following sum of rates
\begin{align}
  \sum_{\vec\outx} c_{\vec \outx} R(\vec\xi \to \vec\outx n), \label{eq:generalsum}
\end{align}
where $\vec \outx=(\outx_1, \outx_2, \cdots, \outx_{n-2})$ is a vector of length $n-2$, with 
$\outx_i$ indicating a photon in mode $i$.  Thus, for two photon in the first three modes
of an interferometer we can have $\vec \outx =(1,1)$, 
$(1,2)$, $(1,3)$, $(2,2)$, $(2,3)$ and $(3,3)$, with $(2,2)$ indicating two photons in mode $2$.  The factor
 $c_{\vec \outx}$ is the inverse of the product of factorials
corresponding to the repetitions in $\vec\outx$. For
instance, for $\vec\outx = (1,2,1,1,2)$, $c_{\vec \outx} = 1/(2!3!)$.  This factor arises because, if a mode contains $k$ photons, 
it must be normalized by dividing the state by $1/\sqrt{k!}$, and the rate by $1/k!$ since the rate is proportional to modulus square of the matrix element involving the state.

Likewise, 
$\vec \xi$ is a vector of length $(n-1)$ where $\xi_i$ has
the same interpretation as $\outx_i$.  To keep the discussion 
simple we assume that all $\xi_i$ are distinct $(c_{\vec \xi}=1)$, although this is not essential.

\par
For $(n-1)$ indistinguishable photons, each rate $R(\vec \xi
\to \vec\outx n)$ is proportional to
the modulus squared for the permanent of the submatrix
$U_{\vec \xi \to \vec\outx n}$, and this permanent is related to the function
\cite{spivak2017immanants}
\begin{align}
&   \sqrt{c_{\vec \outx}} \per (U_{\vec \xi \to \vec\outx n})=\nonumber \\
&\langle (n-1,0,\ldots,0) \vec \outx n \vert {\cal R}_{1\cdots n-1}(\omega_{1\cdots n-1})\bar U
    \vert (n-1,0,\ldots,0) \vec \xi\rangle
\end{align}
where we have again split $U$ into an $\mathrm{SU}(n-1)$ transformation and a coset transformation:
$U= {\cal R}_{1\cdots n-1}(\omega_{1\cdots n-1})\bar U $.  
The strategy is again to insert a complete set of $\mathrm{SU}(n-1)$ states between the subgroup and the 
coset transformations:
\begin{align}
    &\langle (n-1,0,\cdots,0) \vec \outx n \vert {\cal R}_{1\cdots n-1}(\omega_{1\cdots n-1})\bar U
    \vert (n-1,0,\cdots,0) \vec \xi\rangle  \nonumber \\
    &=\sum_{\vec \outy}
    \langle (n-1,0,\cdots,0) \vec \outx n \vert {\cal R}_{1\cdots n-1}(\omega_{1\cdots n-1})
    \vert (n-1,0,\cdots,0)\vec \outy n\rangle \nonumber \\
    &\quad \times \langle (n-1,0,\cdots,0)\vec \outy n \vert \bar U \vert (n-1,0,\ldots,0) \vec \xi\rangle\, ,\\
    &= \sum_{\vec \outy} 
    D^{(n-1,0,\cdots,0)}_{\vec \outx n;\vec \outy n}(\omega_{1\cdots n-1})
    \nonumber \\ &\quad \times 
    \langle (n-1,0,\cdots,0)\vec \outy n \vert \bar U \vert (n-1,0,\ldots,0)\, .
\end{align}
Multiplying by the complex conjugate, summing as per Eq.~\eqref{eq:generalsum}, and using the 
orthogonality of the $D^{(n-1,0,\cdots,0)}_{\vec \outx
n;\vec \outy n}(\omega_{123\cdots n-1})$
functions yields the result.

A similar proof can be
developed for the case of partially distinguishable photons, using the connection between immanants 
of a submatrix of a unitary matrix and group functions.

\section{Concluding remarks}

In this Letter we presented a method of computing sums of coincidence rates using a coset matrix describing a simplified
scattering process, resulting in reduced  computational complexity  compared to the original problem. The result depends on
factoring  the original $n\times n$ scattering matrix into an $\hbox{SU}(n-1)$ matrix 
and a coset matrix, and summing over
states which span
subrepresentations of $\hbox{SU}(n-1)$ inside our many-photon Hilbert space. 

The coset matrices 
$\bar U$ discussed in this Letter
are of the Hessenberg type (though not all Hessenberg
matrices are coset matrices and not every submatrix
of a Hessenberg matrix is Hessenberg), and 
additional simplifications in evaluating permanents of such matrices are possible: we show explicitly that certain sums of 
modulus squared of permanents of $3\times 3$ submatrices of $\hbox{SU}(4)$, can be evaluated using sums of modulus squared of 
\emph{determinants}.    

Additional simplifications in the evaluations of immanants which arise when photons are not all coincident, are 
also known to
occur Hessenberg matrices \cite{Hessenberg}, 
but for the specific case of $3\times 3$ submatrices of $\hbox{SU}(4)$ there is no savings since the  $\Yboxdim{5pt}\yng(2,1)$ is self-conjugate.

We note that good algorithms to evaluate immanants of unitary matrices are difficult to find.  Following Kostant 
\cite{kostan1995immanant} (see also \cite{Dfuntionsandimmanants})  B\"{u}rgisser \cite{burgisser2000computational} proposed to
evaluate immanants using sums of group functions, a strategy that displaces the problem of constructing of such functions.
In  addition, the map $T$ that transforms the evaluation of the immanant of a Hessenberg matrix $\bar U$ to its simplified 
form 
$T(\bar U)$ as per \Eref{eq:per_maps_to_det}, is such that $T(\bar U)$ is not unitary.  As a result the challenge of implementing 
this 
transformation \emph{and} neatly evaluating the simplified immanants by anything other than a brute force method remains an 
open 
problem at this time.

We did not discuss the case where the coset is of the type ${\cal R}\bar U$ with ${\cal R}\in \hbox{SU}(k)$
and $k<n-1$:
the detailed analysis of the possible simplifications arising from the factorization of $\hbox{SU}(k)$ 
submatrices of the original $n\times n$ matrix $U$ remains at
this time an open question.   

When $k$ is small, the savings that result from the summations are small since few $0$s will 
appear
in the coset matrices.  When $k$ is large, the savings are more substantial, although the summations must include rates for 
processes where more than one photon is counted in some detectors.    This suggests that one can devise a series of 
increasingly 
sophisticated tests based on sums to verify the proper functioning of the optical network.  The extent to
which one can construct an efficient witness based on sums of rates remains to be explored, although constructing
coset matrices $\bar U$ from the original $U$ can be done efficiently using Householder transformations 
\cite{urias2010householder,cabrera2010canonical}.



\section{Acknowledgements}
DAA acknowledges support from projects CONACyT 285754, UNAM-PAPIIT IG100518, and from the  Mitacs-CALAREO Globalink Research Award Program.    Dylan~{Spivak} acknowledges support from the Ontario Graduate Scholarship program. 
The work of HdG is supported by NSERC of Canada.  We thank Olivia DiMatteo and Barry Sanders for comments and useful discussions.
\appendix
\section{Immanants of $3\times 3$ matrix}\label{sec:3x3imm}
Immanants are weighted sums of products of matrix entries: 
\begin{align}
    \hbox{Imm}^{\{\lambda\}}(U)=\sum_{\sigma \in S_n} \chi^{\{\lambda\}}(\sigma) 
    U_{\sigma(1)1}U_{\sigma(2)2}\ldots U_{\sigma(n)n}
    \label{eq:immanantdef}
\end{align}
where
$\{\lambda\}$ is a partition of $n$ labelling an irreducible
representation of $S_n$ 
 and $\chi^{\{\lambda\}}(\sigma)$ is the character of $\sigma\in S_n$ for the irrep $\lambda$.  
\par
A convenient mnemonic device to label partitions and therefore irreducible representations of $S_n$ is to use
Young diagrams, 
\cite{band2013quantum,lichtenberg2012unitary,rowe2010fundamentals},\cite{lichtenberg2012unitary}.
A convenient mnemonic device to label partitions and therefore irreducible representations of $S_n$ is to use
Young diagrams
The partition $\{\lambda\}=(\lambda_1,\lambda_2,\ldots,\lambda_k)$ with $\lambda_{k}\ge \lambda_{k+1}$ is pictorially represented by
a left-justified diagram containing $\lambda_k$ boxes on row $k$.  
The partition $\{n\}$ of $n$, used for the permanent, corresponds to the
one-rowed  Young diagram $\Yboxdim{7pt}\yng(3)\cdots \yng(1)$ containing $n$ boxes on the row, while the
partition $\{1^n\}$ used for determinants corresponds to a Young diagram with a single column of $n$ boxes.  


To complete the calculation of an immanant, we need the characters of the appropriate representation.
These can be computed from scratch or found elsewhere\cite{littlewood}.  The characters of
the three irreducible representations of $S_3$ are given in
\tref{table:char}.

\begin{table}[h!]
\centering
{\renewcommand{\arraystretch}{1.5}
\begin{tabular}{|c|c|c|c|c|}
\hline
Elements &$\mathds{1}$&	$\{P_{12},P_{13},P_{23}\}$ & $\{P_{123},P_{132}\}$&	 \\
\hline
irrep $\lambda$ & $\chi^\lambda(\mathds{1})$ & $\chi^\lambda(P_{ab})$\phantom {\Yvcentermath1\Yboxdim{8pt}$\yng(1,1)$} 
&$\chi^\lambda(P_{abc})$ 
& dim. \\
\hline
$\Yvcentermath1\Yboxdim{6pt}\yng(3) $     & 1 & 1 &1 & 1\\
$\Yvcentermath1\Yboxdim{6pt}\yng(2,1) $  & 2 & 0 & -1 & 2 \\
$\Yvcentermath1\Yboxdim{6pt}\yng(1,1,1)$& 1 &-1 &1 &1 \\
\hline
\end{tabular}}
\caption{The character table for $S_3$.} 
\label{table:char}
\end{table}

Using \tref{table:char}, the
immanants for $3\times 3$ matrices are
\begin{align}
    \hbox{Imm}^{\Yboxdim{5pt}\yng(3)}(U)
    &=\sum_\sigma P_\sigma U_{11}U_{22}U_{33} = \hbox{Per}(U)\, ,\\
    \hbox{Imm}^{\Yboxdim{5pt}\yng(1,1,1)}(U)
    &= U_{11}U_{22}U_{33}- (P_{12}+P_{13}+P_{23})U_{11}U_{22}U_{33} \nonumber
  \\&\quad+ (P_{123}+P_{132})U_{11}U_{22}U_{33}= \hbox{Det}(U)\, ,\\
     \hbox{Imm}^{\Yboxdim{5pt}\yng(2,1)}(U)
    &= 2U_{11}U_{22}U_{33}
    - (P_{123}+P_{132})U_{11}U_{22}U_{33}\, ,\nonumber \\
    &=2U_{11}U_{22}U_{33} - U_{12}U_{23}U_{31}-U_{13}U_{21}U_{32}
    \label{eq-immanant-3}
\end{align}
Here, $P_{ijk}$ denotes the
cycle $i\to j\to k\to i$, etc.
Whereas the permanent and the determinant return to themselves to within a sign under permutation of rows or columns, there
is no such simple symmetry for the general immanants or for the $\Yboxdim{7pt}\yng(2,1)$ immanants in particular.  One must construct
linear combinations of these immanants which transform amongst themselves under permutation.

\section{An example of rate calculation}\label{sec:ratecalculation}

Start with $\hat A^\dagger_k(\tau_m)$ as defined in Eq.(\ref{eq:bigAop}).
 We suppose
we have a process with two photons entering ports $1$ and $3$, so the input state is
given by 
\begin{align}
    \ket{\hbox{in}}=\hat A_3^\dagger(\tau_2) \hat A_1^\dagger(\tau_1)\ket{0}\, .
\end{align}
This input state scatters to the output given by
\begin{align}
    \ket{\hbox{out}}&=\int d\frm_1\int d\frm_3
    \phi(\frm_1)\phi(\frm_3)
    e^{-i\tau_1\frm_1}e^{-i\tau_2\frm_3} \nonumber \\
    &\times
    \sum_{p=1}^3a^\dagger_p(\frm_3)U_{p3}\sum_{q=1}^3
    a^\dagger_q(\frm_1)U_{q1}
    \ket{0}\, ,
\end{align}
to be counted in detectors $1$ and $3$ modelled by the product 
\begin{align}
    \hat\Pi_{1,3} &=\hat \Pi_1 \hat \Pi_3\, , \\
    \hat \Pi_k&=  \int d\frM_k a_k^\dagger(\frM_k)
    \ket{0}\bra{0}a_k(\frM_k)
\end{align} 
where $\hat \Pi_k$ models
a flat-spectrum incoherent Fock-number
state measurement operator.  The final coincidence rate given by 
\begin{align}
   & R(13\to 13;\tau_{12})= \bra{\hbox{out}}\,\hat \Pi_{1,3}\ket{\hbox{out}} \nonumber \\
   &\quad =\int d\frM_1
   d\frM_3\,d\tilde\frm_1d\tilde\frm_3\,d\frm_1 d\frm_3\nonumber \\
   &\qquad \times
   \phi^*(\tilde\frm_3)\phi^*(\tilde\frm_1)\phi(\frm_3)\phi(\frm_1)
   e^{-i\tau_1(\frm_1-\tilde\frm_1)}e^{-i\tau_2(\frm_3-\tilde\frm_3)} \nonumber \\
   &\qquad \times \sum_{p'q'}U_{p'3}^\dagger U_{q'1}^\dagger 
   \bra{0} a_{q'}(\tilde \frm_1)a_{p'}(\tilde \frm_3)
   a^\dagger_3(\frM_3)
   a^\dagger_1(\frM_1)\ket{0}\, \nonumber \\
   &\qquad \times \sum_{pq}U_{p3}U_{q1} 
   \bra{0} a_1(\frM_1)a_{3}(\frM_3) a^\dagger_p(\frm_3)
   a^\dagger_1(\frm_1)\ket{0}\, .
\end{align}
Using now the boson commutation relations
\begin{align}
    [a_k^\dagger(\frm_\outa),a_m(\frm_\outb)]=-\delta_{km}
    \delta(\frm_\outa-\frm_\outb)
\end{align}
we have
\begin{align}
    &a_1(\frM_1)a_{3}(\frM_3) a^\dagger_p(\frm_3)
   a^\dagger_1(\frm_1)\ket{0}\nonumber \\
   &=\delta_{p3}\delta(\frM_3-\frm_3)\delta_{q1}\delta(\frM_1-\frm_1)+
   \delta_{p1}\delta(\frM_1-\frm_3)\delta_{q3}\delta(\frM_3-\frm_1)\, .
\end{align}
The rate then becomes
\begin{align}
  & R(13\to 13;\tau_{12}) =\int d\frM_1
  d\frM_3\,d\tilde\frm_1d\tilde\frm_3\,d\frm_1 d\frm_3\nonumber \\
   &\qquad \times
   \phi^*(\tilde\frm_3)\phi^*(\tilde\frm_1)\phi(\frm_3)\phi(\frm_1)
   e^{-i\tau_1(\frm_1-\tilde\frm_1)}e^{-i\tau_2(\frm_3-\tilde\frm_3)} \nonumber \\
   &\qquad\times \sum_{p'q'}U^\dagger_{p'3}U^\dagger_{q'1}
    \bra{0} a_{q'}(\tilde \frm_1)a_{p'}(\tilde \frm_3) a^\dagger_3(\frM_3)
   a^\dagger_1(\frM_1)\ket{0}\nonumber \\
   &\qquad\times
   \left[U_{11}U_{33}\delta(\frM_1-\frm_1)\delta(\frM_3-\frm_3) 
   \right.\nonumber \\
 &\qquad\qquad \left.   +U_{13}U_{31}\delta(\frM_1-\frm_3)\delta(\frM_3-\frm_1)\right]\, .
\end{align}
For economy it is convenient to write
\begin{align}
  &U_{11}U_{33}\delta(\frM_1-\frm_1)\delta(\frM_3-\frm_3) 
   \nonumber \\
 &\qquad\qquad  +U_{13}U_{31}\delta(\frM_1-\frm_3)\delta(\frM_3-\frm_1)\nonumber \\  
 &=\sum_{\sigma=\mathbbm{1},P_{13}}U_{1\sigma(1)}U_{3\sigma(3)}\delta(\frM_1-\frm_{\sigma(1)})\delta(\frM_3-\frm_{\sigma(3)})\, .
\end{align}
Using again the commutation relations to evaluate the expectation value
$\bra{0} a_{q'}(\tilde \frm_1)a_{p'}(\tilde \frm_3) a^\dagger_3(\frM_3)
   a^\dagger_1(\frM_1)\ket{0}$ we obtain this time
   \begin{align}
  & R(13\to 13;\tau_{12}) =\int d\frM_1 d\frM_3\,d\tilde\frm_1d\tilde\frm_3\,d\frm_1 d\frm_3\nonumber \\
   &\qquad \times \phi^*(\tilde\frm_3)\phi^*(\tilde\frm_1)\phi(\frm_3)\phi(\frm_1)
   e^{-i\tau_1(\frm_1-\tilde\frm_1)}e^{-i\tau_2(\frm_3-\tilde\frm_3)} \nonumber \\
   &\times 
   \left(\sum_{\sigma=\mathbbm{1},P_{13}}
   U^\dagger_{1\sigma(1)}U^\dagger_{3\sigma(3)}\delta(\tilde\frm_1-\frM_{\sigma(1)})
   \delta(\tilde\frm_3-\frM_{\sigma(3)})
  \right)
    \nonumber \\
   &\times 
\left(\sum_{\sigma=\mathbbm{1},P_{13}}U_{1\sigma(1)}U_{3\sigma(3)}\delta(\frM_1-\frm_{\sigma(1)})
\delta(\frM_3-\frm_{\sigma(3)})\right)\, .
\end{align}
Assuming now for simplicity 
\begin{align}
  \vert\phi(\frm_k)\vert^2=\frac{e^{-(\frm_k-\frm_0)^2/2\mathrm{s}^2}}{\sqrt{2\pi}\mathrm{s}}
\end{align}
we obtain the final result 
\begin{align}
    &R(13\to 13;\tau_{12})=\vert U_{11}U_{33}\vert^2+ \vert U_{13}U_{31}\vert^2\nonumber \\
    &\qquad +e^{-\mathrm{s}^2\tau_{12}^2}\left(U^\dagger_{11}U^\dagger_{33}U_{31}U_{13}+ U_{11}U_{33}
    U^\dagger_{31}U^\dagger_{13}\right)\, ,\\
    &=\textstyle\frac{1}{2}\left(1+e^{-\mathrm{s}^2\tau_{12}^2}\right)\vert\hbox{Per}(U)\vert^2
    +\textstyle\frac{1}{2}\left(1-e^{-\mathrm{s}^2\tau_{12}^2}\right)\vert\hbox{Det}(U)\vert^2\, .
\end{align}




\bibliography{mybibfile}
\end{document}